\documentclass{article}

\usepackage[a4paper,top=2.5cm,bottom=2.5cm,left=2.5cm,right=2.5cm,marginparwidth=2cm]{geometry}
\usepackage{microtype}
\usepackage{chicago}
\usepackage{graphicx}
\usepackage{subfigure}
\usepackage{booktabs} 

\usepackage{amssymb,amsmath,amsfonts,amscd,amstext,bbm, bm, enumerate, dsfont, mathtools}
\usepackage{enumitem}
\usepackage{graphicx}
\usepackage{microtype}
\usepackage{epstopdf}
\usepackage{booktabs}
\usepackage{algorithm,algpseudocode}
\usepackage{verbatim}
\usepackage{tcolorbox}
\usepackage{import}
\usepackage{tikz,pgfplots}
\usepackage{subfig}
\usetikzlibrary{tikzmark,shapes}
\usetikzlibrary{matrix}
\usepgfplotslibrary{groupplots}
\pgfplotsset{compat=newest}
\usepackage{tabularx}
\usepackage{soul}
\usepackage{array}
\usepackage{balance}
\usepackage{tabularx}
\usepgfplotslibrary{statistics}
\ifpdf
  \DeclareGraphicsExtensions{.eps,.pdf,.png,.jpg}
\else
  \DeclareGraphicsExtensions{.eps}
\fi

\newtheorem{theorem}{Theorem}
\newtheorem{lemma}{Lemma}

\newtheorem{assumption}{Assumption}

\newenvironment{Proof}{\par\noindent{\bf Proof.}}{\hfill$\square$}
\usepackage{multicol}
\usepackage{multirow}
\usepackage{url}
\usepackage{bbold}

\newcommand{\w}{\boldsymbol{\omega}}

\newcommand{\bloc}{\mathcal B}

\DeclareMathOperator*{\argmin}{arg\,min}
\DeclareMathOperator{\Var}{Var}

\newcommand{\Loss}{\mathcal{L}}

\newcommand{\userS}{\mathcal{U}}
\newcommand{\itemS}{\mathcal{I}}
\newcommand{\vecU}{\mathbf{U}}
\newcommand{\vecI}{\mathbf{V}}
\newcommand{\prefu}{\renewcommand\arraystretch{.2} \begin{array}{c}
   {\succ} \\  \mbox{{\tiny {\it u}}}
  \end{array}\renewcommand\arraystretch{1ex}}
\newcommand{\posI}{\mathcal{I}^{+}}
\newcommand{\negI}{\mathcal{I}^{-}}

\newcommand{\MovieL}{\textsc{MovieLens}}
\newcommand{\NetF}{\textsc{Netflix}}

\newcommand{\Out}{\textsc{Outbrain}}
\newcommand{\ML}{\textsc{ML}}
\newcommand{\R}{\mathbb{R}}
\newcommand{\kasandr}{\textsc{Kasandr}}
\newcommand{\PANDOR}{\textsc{Pandor}}

\newcommand{\MostPop}{\texttt{MostPop}}

\newcommand{\MF}{\texttt{MF}}
\newcommand{\GRU}{\texttt{GRU4Rec}}
\newcommand{\ProdVec}{\texttt{Prod2Vec}}
\newcommand{\BPR}{\texttt{BPR}}

\newcommand{\SO}{\texttt{SAROS}}
\newcommand{\batch}{\texttt{BPR$_b$}}

\newcommand{\mapk}{\texttt{MAP@K}}
\newcommand{\apk}{\texttt{AP@K}}
\newcommand{\mapfive}{\texttt{MAP@5}}
\newcommand{\mapten}{\texttt{MAP@10}}
\newcommand{\ndcgfive}{\texttt{NDCG@5}}
\newcommand{\ndcgten}{\texttt{NDCG@10}}
\newcommand{\ndcgk}{\texttt{NDCG@K}}
\newcommand{\dcgk}{\texttt{DCG@K}}
\newcommand{\idcgk}{\texttt{IDCG@K}}
\renewcommand{\w}{\boldsymbol{\omega}}

\usepackage[T1]{fontenc}
\usepackage[utf8]{inputenc}
\usepackage{authblk}

\title{Sequential Learning over Implicit Feedback for Robust Large-Scale Recommender Systems}
\author[1]{Alexandra Burashnikova \thanks{Corresponding author, aleksandra.burashnikova@skoltech.ru}}
\author[1,2]{Yury Maximov}
\author[3]{Massih-Reza Amini}
\affil[1]{Skolkovo Institute of Science and Technology}
\affil[2]{Theoretical Division T-5 and CNLS, Los Alamos National Laboratory}
\affil[3]{University Grenoble-Alpes}

\date{\today}

\begin{document}

\maketitle
\begin{abstract}
In this paper, we propose a robust sequential learning strategy for training large-scale Recommender Systems (RS) over implicit feedback mainly in the form of clicks. 
Our approach relies on the minimization of a pairwise ranking loss over blocks of consecutive items constituted by a sequence of non-clicked items followed by a clicked one for each user. 
Parameter updates are discarded if for a given user the number of sequential blocks is below or above some given thresholds estimated over the distribution of the number of blocks in the training set.
This is to prevent from an abnormal number of clicks over some targeted items, mainly due to bots; or very few user interactions. Both scenarios affect the decision of RS and imply a shift over the distribution of items that are shown to the users. We provide a theoretical analysis showing that in the case where the ranking loss is convex, the deviation between the loss with respect to the sequence of weights found by the proposed algorithm and its minimum is bounded.  Furthermore, experimental results on five large-scale collections demonstrate the efficiency of the proposed algorithm with respect to the state-of-the-art approaches, both regarding different ranking measures and computation time.
\end{abstract}

\section{Introduction}
\label{Introduction}

With the increasing number of products available online, there is a surge of interest in the design of automatic systems --- generally referred to as Recommender Systems (RS) --- that provide personalized recommendations to users by adapting to their taste. The study of RS has become an active area of research these past years, especially since the  Netflix Price \cite{Bennett:07}. 

One characteristic of online recommendation is the huge unbalance between the available number of products and those shown to the users. Another aspect is the existence of bots that interact with the system by providing too many feedback over some targeted items; or many users that do not interact with the system over the items that are shown to them. In this context, the main challenges concern the design of a scalable and an efficient online RS in the presence of noise and unbalanced data, and they have evolved over time with the continuous development of data collections released for competitions or issued from e-commerce\footnote{\scriptsize \url{https://www.kaggle.com/c/outbrain-click-prediction}}.  New approaches for RS now primarily consider {\it implicit} feedback, mostly in the form of clicks, that are easier to collect than {\it explicit} feedback which is in the form of scores. Implicit feedback is more challenging to deal with as they do not clearly depict the preference of a user over items, i.e., (no)click does not necessarily mean (dis)like \cite{Hu:2008}. For this case, most of the developed approaches are based on the Learning-to-rank paradigm \cite{Liu:2009} and focus on how to leverage the click information over the unclick one without considering the sequences of users' interactions. 

In this paper, we propose a SequentiAl RecOmmender System for implicit feedback (called \SO), that updates the model parameters user per user over blocks of items constituted by a sequence of unclicked items followed by a clicked one. The parameter updates are discarded for users who interact very little or a lot with the system. For other users, the update is done by minimizing the average ranking loss of the current model that scores the clicked item below the unclicked ones in a corresponding block. Other approaches to modeling the sequences of users feedback begin to raise, but they suffer from a lack of theoretical analysis formalizing the overall learning strategy. 
In this work, we analyze the convergence property of the proposed approach and show that in the case where the global ranking loss estimated over all users and items is convex; then the minimizer found by the proposed sequential approach converges to the minimizer of the global ranking loss. 

Experimental results conducted on five large publicly available datasets show that our approach is highly competitive compared to the state-of-the-art models and, it is significantly faster than both the batch and the online versions of the algorithm which, under some instantiation bear similarity with the Bayesian Personalized Ranking model \cite{rendle_09}. 

The rest of this paper is organized as follows. Section \ref{sec:soa} relates our work to previously proposed approaches. Section \ref{sec:Frame} introduces the general ranking learning problem that we address in this study. Then, in Section~\ref{sec:TA}, we present the \SO{} algorithm and provide an analysis of its convergence. Section \ref{sec:Exps} presents the experimental results that support this approach. Finally, in Section \ref{sec:Conclusion}, we discuss the outcomes of this study and give some pointers to further research.

\section{Related work}
\label{sec:soa}

Two main approaches have been proposed for recommender systems. The first one, referred to as Content-Based recommendation or cognitive filtering \cite{Pazzani2007}, makes use of existing contextual information about the users (e.g., demographic information) or items (e.g., textual description) for the recommendation. The second approach referred to as Collaborative Filtering and undoubtedly the most popular one \cite{Su:2009}, relies on past interactions and recommends items to users based on the feedback provided by other similar users. 

Traditionally, collaborative filtering systems were designed using  {\it explicit} feedback, mostly in the form of rating \cite{Koren08}. However,  rating information is non-existent on most e-commerce websites and is challenging to collect, and user interactions are often done sequentially. Recent RS systems focus on learning scoring functions using {\it implicit} feedback, in order to assign higher scores to clicked items than to unclicked ones rather than to predict the clicks as it is usually the case when we are dealing with explicit feedback \cite{Cremonesi:2010,He2016,Pessiot:07,rendle_09,Volkovs2015,Zhang:16}.  

The main idea here is that even a clicked item does not necessarily express the preference of a user for that item, it has much more value than a set of unclicked items for which no action has been made. In most of these approaches, the objective is to rank the clicked item higher than the unclicked ones by finding a suitable representation of users and items in a way that for each user the ordering of the clicked items over unclicked ones is respected by dot product in the joint learned space.

One common characteristic of publicly available collections for recommendation systems is the huge unbalance between positive (click) and negative feedback (no-click) in the set of items displayed to the users, making the design of an efficient online RS extremely challenging. To deal with this problem; some works propose to weight the impact of positive and negative feedback directly in the objective function \cite{Hu:2008,Pan:2008} or to sample the data over a predefined buffer before learning \cite{Liu2016}, but these approaches do not model the shift over the distribution of positive and negative items, and results on new test data may be affected. Other approaches tackle the sequential learning problem for RS by taking into account the temporal aspect of interactions directly in the design of a dedicated model and are mainly based on Markov Models (MM), Reinforcement Learning (RL) and Recurrent Neural Networks (RNN) \cite{Liu:2009,Donkers:2017}. Recommender systems based on Markov Models, consider the sequential interaction of users as a stochastic process over discrete random variables related to predefined user behavior. These approaches suffer from some limitations mainly due to the sparsity of the data leading to a poor estimation of the transition matrix \cite{GuyShani,HeQi,garcin2013,Haidong,Sahoo}. 

Various strategies have been proposed to leverage the impact of sparse data, for example by considering only the last frequent sequences of items and using finite mixture models \cite{GuyShani}, or by combining similarity-based methods with high-order Markov Chains \cite{He,Sahoo}. Although it has been shown that in some cases the proposed approaches can capture the temporal aspect of user interactions but these models suffer from high complexity and generally they do not pass the scale.  Some other approaches consider RS as a Markov decision process (MDP) problem and solve it using reinforcement learning (RL) \cite{Moling,Tavakol}. The size of discrete actions bringing the RL solver to a larger class of problems is also a bottleneck for these approaches. Very recently Recurrent neural networks such as GRU or LSTM, have been proposed for personalized recommendations \cite{JiaLi,hidasi,hidasi2,hidasi3},  where the input of the network is generally the current state of the session, and the output is the predicted preference over items (probabilities for each item to be clicked next).

Our proposed strategy differs from other sequential based approaches in the way that the model parameters are updated, at each time a block of unclicked items followed by a clicked one is constituted; and by controlling the number of blocks per user interaction. If for a given user, this number is below or above two predefined thresholds found over the distribution of the number  of block,  parameter updates for that particular user are discarded. We prove that in the case where the general ranking loss over all users is convex;  the minimizer found by the proposed sequential algorithm converges to the true minimizer of this loss.

\section{Problem Setting and Framework}
\label{sec:Frame}
Throughout, we use the following notation. For any positive integer $n$, $[n]$ denotes the set $[n]\doteq \{1,\ldots,n\}$. We suppose that $\itemS\doteq [M]$ and $\userS\doteq [N]$ are two sets of indexes defined over items and users. Further, we assume that a pair constituted by a user $u$ and an item $i$ is identically and independently distributed according to a fixed yet unknown distribution ${\cal D}_{\cal U, \cal I}$. 

At the end of his or her session, a user $u\in\userS$ has reviewed a subset of items $\itemS_u\subseteq \itemS$ that can be decomposed into two sets: the set of preferred and non-preferred items denoted by $\posI_u$ and $\negI_u$, respectively. Hence, for each pair of items $(i,i')\in\posI_u\times \negI_u$, the user $u$ prefers item $i$ over item $i'$; symbolized by the relation $i\!\prefu\! i'$. From this preference relation a desired output $y_{u,i,i'}\in\{-1,+1\}$ is defined over the pairs $(u,i)\in\userS\times\itemS$ and $(u,i')\in\userS\times\itemS$, such that $y_{u,i,i'}=+1$ if and only if   $i\!\prefu\! i'$.  We suppose that the indexes of users in as well as those of items in the set $\itemS_u$, shown to the active user $u\in\userS$, are ordered by time. 

Finally, for each user $u$, parameter updates are performed over blocks of consecutive items where a  block $\bloc_u^\ell=\text{N}_u^{\ell}\sqcup\Pi_u^{\ell}$, corresponds to a (time-ordered) sequence of no-preferred items, $\text{N}_u^{\ell}$, and at least one preferred one, $\Pi_u^{\ell}$. Hence, $\posI_u=\bigcup_\ell \Pi_u^{\ell}$ and $\negI_u=\bigcup_\ell \text{N}_u^{\ell}; \forall u\in\userS$.

\subsection{Learning Objective}
\label{sec:LO}
Our objective here is to minimize an expected error penalizing the misordering of all pairs of interacted items $i$ and $i'$ for a user $u$. Commonly, this objective is given under the Empirical Risk Minimization principle \cite{Vapnik2000} by minimizing the empirical ranking loss estimated over the items and the final set of users who interacted with the system~:
\begin{equation}
\label{eq:RL}
\!\widehat{\Loss}_u(\w)\!=\!\frac{1}{|\posI_u||\negI_u|}\!\sum_{i\in \posI_u}\!\sum_{i'\in \negI_u} \!\ell_{u,i,i'} (\w),
\end{equation}
and $\Loss(\w) = {\mathbb E}_{u} \left[\widehat{\cal L}_u(\w)\right]$, where ${\mathbb E}_{u}$ is the expectation with respect to users chosen randomly according to the uniform distribution, and  $\widehat{\Loss}_u(\w)$ is the pairwise ranking loss with respect to user $u$'s interactions. As in other studies, we represent each user $u$ and each item $i$ respectively by vectors $\vecU_u\in\R^k$ and $\vecI_i\in\R^k$ in the same latent space of dimension $k$ \cite{Koren:2009}. The set of weights to be found $\w=(\vecU,\vecI)$, are then matrices formed by the vector representations  of users $\vecU=(\vecU_u)_{u\in [N]}\in\R^{N\times k}$ and items $\vecI=(\vecI_i)_{i\in[M]}\in\R^{M\times k}$. 
The minimization of the ranking loss above in the batch mode with the goal of finding user and item embeddings, such that the dot product between these representations in the latent space reflects the best the preference of users over items, is a common approach. Other strategies have been proposed for the minimization of the empirical loss \eqref{eq:RL}, among which the most popular one is perhaps the Bayesian Personalized Ranking (\BPR) model \cite{rendle_09}. In this approach, the instantaneous loss, $\ell_{u,i,i'}$, is the surrogate regularized logistic  loss~:
\begin{align}
\hspace{-1mm}\ell_{u,i,i'}(\w) = & \log\left(1+e^{-y_{i,u,i'}\vecU_u^\top(\vecI_{i}-\vecI_{i'})}\right) \nonumber \\ 
& + \mu (\|\vecU_u\|_2^2+\|\vecI_{i}\|_2^2 + \|\vecI_{i'}\|_2^2),  \mu \ge 0. \label{eq:instloss}
\end{align}

The {\BPR} algorithm proceeds by first randomly choosing a user $u$, and then repeatedly selecting two pairs $(i,i')\in \itemS_u\times \itemS_u$. In the case where one of the chosen items is preferred over the other one (i.e. $y_{u,i,i'}\in\{-1,+1\}$), the algorithm then updates the weights using the stochastic gradient descent method over the instantaneous loss \eqref{eq:instloss}.  In this case, the expected number of rejected pairs is proportional to  $O(|\itemS_u|^2)$ \cite{Sculley09largescale} which may be time-consuming in general. Another drawback is that user preference over items depend mostly on the context where these items are shown to the user. A user may prefer (or not) two items independently one from another, but within a given set of shown items, he or she may completely have a different preference over these items. By sampling items over the whole set of shown items, this effect of local preference is generally undermined. 

Another particularity of online recommendation that is not explicitly taken into account by existing approaches is the bot attacks in the form of excessive clicks over some target items.  They are made to force the RS to adapt its recommendations toward these target items, or a very few interactions which in both cases introduce biased data for the learning of an efficient RS. 

In order to tackle these points, our approach updates the parameters whenever the number of constituted blocks per user is lower and upper-bounded. In this case, at each time  a block $\bloc_u^\ell=\text{N}_u^{\ell}\sqcup\Pi_u^{\ell}$ is formed; weights are updated by miniminzing the ranking loss corresponding to this block~:
\begin{equation}
\label{eq:CLoss}
{\widehat {\cal L}}_{\bloc_u^\ell}(\w_u^{\ell}) = \frac{1}{|\Pi_u^{\ell}||\text{N}_u^{\ell}|}\sum_{i \in \Pi_u^{\ell}} \sum_{i'\in \text{N}_u^{\ell}} \ell_{u, i, i'} ({\w}_u^{\ell}).
\end{equation}

\begin{figure}[!t]
    \centering
  \hspace{-3mm}  \includegraphics[width=\textwidth]{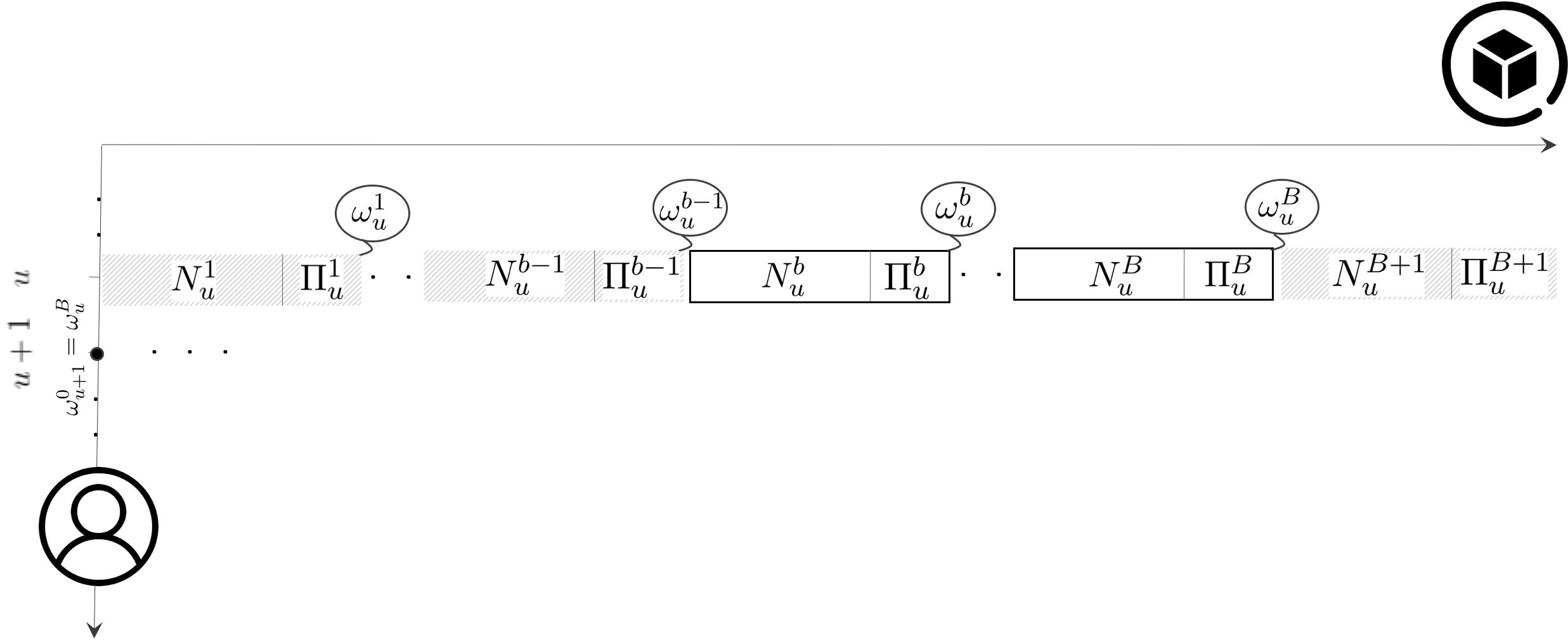}
    \caption{A pictorial depiction of the sequential updates of weights $(\omega_u^{\ell})_{1\leq \ell\leq B}$ for a user $u\in\userS$. The horizontal axis represents the sequence of interactions over items ordered by time. Each update of weights $\omega_u^{\ell}; \ell\in\{b,\ldots,B\}$ occurs whenever the corresponding sets of negative interactions, $\text{N}^\ell_u$, and positive ones, $\Pi_u^\ell$, exist; and whenever $\ell$ is lower and upper-bounded by respectively the minimum and maximum allowed interactions, $b$ and $B$. For a new user $u+1$, the initial weights $\omega_{u+1}^0=\omega_u^B$ are the ones obtained from the last update of the previous user's interactions.}  
    \label{fig:SILICOM}
\end{figure}

This procedure is depicted in Figure \ref{fig:SILICOM}.

\subsection{Algorithm \SO{}}
\label{sec:Algo}



\begin{algorithm}[!ht]
   \caption*{{\bf Algorithm} {\SO}: SequentiAl RecOmmender System}
\begin{algorithmic}
   \State {\bfseries Input:} A time-ordered sequence (user and items)  $\{(u,(i_1, \dots, i_{|\itemS_u|})\}_{u=1}^N$ drawn i.i.d. from~${\cal D}_{{\cal U}, {\cal I}}$
   \State {\bfseries Input:} maximal $B$ number of blocks allowed per user $u$
   \State {\bfseries Input:} initial parameters $\omega_1^0$, and (possibly non-convex) surrogate loss function $\ell(\omega)$
   \For{$u \in\userS$} 
        \State Let $\text{N}_u^{\ell} = \varnothing, \, \Pi_u^\ell = \varnothing$ be the sets of positive and negative items, iteration counter $\ell = 0$
        \For{$\ell\leq I_u$ and $\ell \leq B$}
            \If{ $u$ \text{provides a negative feedback on }$i_k$}
                \State $\text{N}_u^\ell \leftarrow \text{N}_u^\ell \cup \{i\}$
            \Else
            \State $\Pi_u^\ell \leftarrow \Pi_u^\ell \cup \{i\}$
            \EndIf
            \If{ $\text{N}_u^{\ell} \neq \varnothing$ and $\Pi_u^{\ell} \neq \varnothing$ and $\ell
            \leq B$}
                \State {\small $\w_u^{\ell+1} \leftarrow \w_u^\ell - \frac{\eta}{|{\text{N}_u^{\ell}}||\Pi_u^{\ell}|}  \displaystyle{\sum_{i\in\Pi_u^{\ell}}\sum_{i'\in \text{N}_u^{\ell}}} \nabla \ell_{u,i,i'} (\w_{u}^\ell)$}
                \State $\ell = \ell+1, \text{N}_u^\ell = \varnothing$, $\Pi_u^\ell = \varnothing$
            \EndIf
        \EndFor
        \State ${\omega}_{u+1}^0 = {\omega}_{u}^\ell$
   \EndFor
   \State {\bfseries Return:} 
    ${\bar\omega}_N = \sum_{u\in\userS} \omega_u^0$ 
\end{algorithmic}
\end{algorithm}

The pseudo-code of {\SO} is shown in the following. Starting from initial weights $\w_1^0$ chosen randomly for the first user. For each current user $u$, having been shown $I_u$ items, the sequential update rule consists in updating the weights block by block where after $\ell$ updates; where the $(\ell+1)^{th}$ update over the current block $\bloc_u^\ell=\text{N}_u^{\ell}\sqcup\Pi_u^{\ell}$ corresponds to one gradient descent step over the ranking loss estimated on these sets and which with the current weights $\w_u^{\ell}$ writes,

To prevent from a very few interactions or from bot attacks, two thresholds $b$ and $B$ are fixed over the parameter updates. For a new user $u+1$, the parameters are initialized as the last updated weights from the previous user's interactions in the case where the corresponding number of updates $\ell$ was in the interval $[b,B]$; i.e. $\omega^0_{u+1}=\omega^\ell_u$. On the contrary case, they are set to the same previous initial parameters; i.e., $\omega^0_{u+1}=\omega^0_u$.

\section{Analysis}
\label{sec:TA}

We provide proofs of convergence for the {\SO} algorithm under the typical hypothesis that the system is not instantaneously affected by the sequential learning of the weights. This hypothesis corresponds to the generation of items shown to users independently and identically distributed with some stationary in time underlying distribution~${\cal D}_{\cal I}$ and constitutes the main hypothesis of almost all the existing studies. 

We conduct our analysis under the following technical Assumption~:

\begin{assumption}\label{asmp:smooth}
Let the loss functions $\ell_{u, i, i'}(\omega)$ and ${\cal L}(\omega)$, $\omega\in\mathbb{R}^d$ be such that for some absolute constants $\gamma\ge \beta > 0$ and $\sigma>0$~:
\begin{enumerate}
    \item $\ell_{u, i, i'}(\omega)$ is non-negative for any user and a pair of items $(u, i, i')$;
    \item $\ell_{u, i, i'}(\omega)$ is twice continuously differentiable, and for any user $u$ and a pair of items $(i,i')$ \begin{align*}
        \gamma \|\omega - \omega'\|_2 \ge
        &
        \|\nabla \ell_{u, i, i'}(\omega) - \nabla \ell_{u, i, i'}(\omega')\|_2 , 
        \\
        \beta \|\omega - \omega'\|_2 \ge & \|\nabla {\cal L}(\omega) -  \nabla
        {\cal L}(\omega')\|_2, 
    \end{align*}
    \item Variance of the empirical loss is bounded
    \[
        \mathbb{E}_{\cal D}\left\|\nabla \widehat{\cal L}_u(\omega) - \nabla {\cal L}(\omega)\right\|_2^2 \le \sigma^2. 
    \]
\end{enumerate}
Furthermore, there exist some positive lower and upper bounds $b$ and $B$, such that the number of updates for any $u$ is within the interval $[b, B]$ almost surely.
\end{assumption}

Our main result is the following theorem which provides a bound over the deviation of the  ranking loss  with respect to the sequence of weights found by the {\SO} algorithm and its minimum in the case where the latter is convex.

\begin{theorem}\label{thm:10}
Let $\ell_{u, i, i'}(\omega)$ and ${\cal L}(\omega)$ satisfy Assumption~\ref{asmp:smooth}.
Then for any constant step size $\eta$, verifying $0< \eta \le \frac{1}{\beta B}$, $0< \eta \le 1/\sqrt{UB(\sigma^2 + 3\gamma^2/b)}$, and any set of users $\userS\doteq [U]$; algorithm {\SO} iteratively generates  a sequence $\{\omega_j^0\}_{u\in \userS}$ such that 
\begin{gather*}
    \frac{1}{\beta} \mathbb{E}\|\nabla {\cal L}({\omega}_u^0)\|_2^2
    \le
    \frac{\beta B \Delta_{\cal L}^2}{u} + 2\Delta_{\cal L}\sqrt{\frac{B\sigma^2 + 3B\gamma^2/b}{u}}, 
\end{gather*}
where $\Delta_{\cal L}^2 = {\frac{2}{\beta}({\cal L}(\omega_0) - {\cal L}(\omega^*))}$, and the expectation is taken with respect to users chosen randomly according to the uniform distribution $p_u=\frac{1}{N}$.

Furthermore, if the ranking loss ${\cal L}(\omega)$ is convex, then for the sequence $\{\omega_j^0\}_{u\in \userS}$ generated by algorithm {\SO} and $\bar\omega_u = \sum_{j\le u} \omega_j^0$ we have
\begin{gather*}
    {\cal L}({\bar \omega}_u) - {\cal L}({\omega_*}) \le \frac{\beta B \Delta_\omega^2}{u} + 2\Delta_{\omega}\sqrt{\frac{B\sigma^2 + 3B\gamma^2/b}{u}},
\end{gather*}
where $\Delta_{\omega} = \|\omega_0 - \omega_*\|_2^2$, and $\omega_*=\mathop{\argmin}_\omega {\cal L}({\omega})$.
\end{theorem}

The proof is provided in the Supplementary, and it is based on the earlier result of \cite{ghadimi2013stochastic} for the randomized stochastic gradient descent. This result implies that the loss over a sequence of weights $(\omega_j^0)_{u\in \userS}$ generated by the algorithm converges to the true minimizer of the ranking loss ${\cal L}(\omega)$ with a rate proportional to $O(\frac{1}{\sqrt{u}})$. The stochastic gradient descent strategy implemented in the Bayesian Personalized Ranking model (\BPR) \cite{rendle_09} also converges to the minimizer of the ranking loss ${\cal L}(\omega)$ with the same rate. However, the main difference between \BPR{} and \SO{} is their computation time. As stated in section \ref{sec:LO} the expected number of rejected random pairs sampled by algorithm \BPR{} before making one update is $O(|\itemS_u|^2)$ while with \SO{}, blocks are created sequentially as and when users interact with the system.  For each user $u$, weights are updated whenever a block is created, with the overall complexity of   $O(\max_\ell (\Pi_u^\ell||\times |\text{N}_u^\ell|))$, with $\max_\ell (|\Pi_u^\ell|\times |\text{N}_u^\ell|)\ll|\itemS_u|^2$.

\section{Experimental Setup and Results}
\label{sec:Exps}

In this section, we provide an empirical evaluation of our optimization strategy on some popular benchmarks proposed for evaluating RS. All subsequently discussed components were implemented in Python3 using the TensorFlow library.\footnote{\url{https://www.tensorflow.org/}.}
We first proceed with a presentation of the general experimental set-up, including a description of the datasets and the baseline models.

\paragraph{Datasets. } We report results obtained on five publicly available data\-sets, for the task of personalized Top-N recommendation on the following collections~:
\begin{itemize}
\item \ML-1M \cite{Harper:2015:MDH:2866565.2827872} and {\NetF}\footnote{\url{http://academictorrents.com/details/9b13183dc4d60676b773c9e2cd6de5e5542cee9a}} consist of user-movie ratings, on a scale of one to five, collected from a movie recommendation service and the Netflix company. The latter was released to support the Netflix Prize competition\footnote{B. James and L. Stan, The Netflix Prize (2007).}. \ML-1M dataset gathers 1,000,000 ratings and {\NetF} consists of 100 million ratings. For both datasets, we consider ratings greater or equal to $4$ as positive feedback, and negative feedback otherwise.
\item 
We extracted a subset out of the {\Out} dataset from of the Kaggle challenge\footnote{\url{https://www.kaggle.com/c/outbrain-click-prediction}} that consisted in the recommendation of news content to users based on the 1,597,426 implicit feedback collected from multiple publisher sites in the United States.
\vspace{1mm}\item  {\kasandr}\footnote{\url{https://archive.ics.uci.edu/ml/datasets/KASANDR}} dataset \cite{sidana17}  
contains 15,844,717 interactions of 2,158,859 users in Germany using Kelkoo's (\url{http://www.kelkoo.fr/}) online advertising platform.
\item {\PANDOR}\footnote{\url{https://archive.ics.uci.edu/ml/datasets/PANDOR}} is another publicly available dataset for online recommendation \cite{sidana18}  provided by Purch (\url{http://www.purch.com/}). The dataset records 2,073,379 clicks generated by 177,366 users of one of the Purch's high-tech website over 9,077 ads they have been shown during one month. 
\end{itemize}

\begin{table}[!t]
    \centering
    \begin{tabular}{llllll}
    \hline
    Data&$|\mathcal{U}|$&$|\mathcal{I}|$&Sparsity&Avg. \# of $+$ & \!\!\!\!Avg. \# of $-$\\
    \hline
    {\ML}-1M&6,040&3,706&.9553&95.2767& \!\!\!\!70.4690\\
   \Out&49,615&105,176&.9997&6.1587& \!\!\!\!26.0377 \\
   \PANDOR&177,366&9,077&.9987&1.3266& \!\!\!\!10.3632\\
     \NetF&90,137&3,560&.9914&26.1872& \!\!\!\!20.2765\\
   \kasandr&2,158,859&291,485&.9999&2.4202& \!\!\!\!51.9384\\
        \hline
    \end{tabular}
    
    \caption{Statistics on the \# of users and items; as well as the sparsity and the average number of $+$ (preferred) and $-$ (non-preferred) items on {\ML}-1M, {\NetF}, {\Out}, {\kasandr} and {\PANDOR} collections after preprocessing considered in our experiments.}
    \label{tab:datasets}
\end{table}

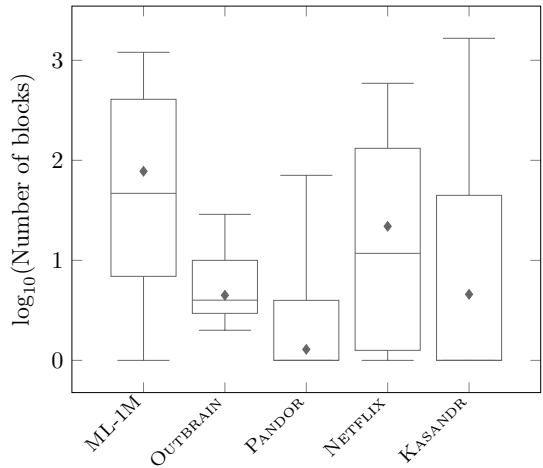
\begin{figure}
    \begin{center}
    \begin{tikzpicture}[scale=0.9]
  \begin{axis}
    [ 
    ylabel = {$\log_{10}(\text{Number of blocks})$},
    boxplot/draw direction=y,
    xtick={1,2,3,4,5},
    xticklabels={\ML-1M, \Out,\PANDOR,\NetF,\kasandr},
    x tick label style={font=\footnotesize, rotate=45,anchor=east}
    ]
    \addplot+[mark = *, mark options = {black!30},
    boxplot prepared={
      lower whisker=0,
      lower quartile=0.84,
      median=1.67,
      average = 1.89,
      upper quartile=2.61,
      upper whisker=3.08
    }, color = black!60
    ]coordinates{};
    \addplot+[mark = *, mark options = {black!30},
    boxplot prepared={
      lower whisker=0.301,
      lower quartile=0.47,
      median=0.602,
      average = 0.651,
      upper quartile=1,
      upper whisker=1.46
    }, color = black!60
    ]coordinates{};
       \addplot+[mark = *, mark options = {black!30},
    boxplot prepared={
      lower whisker=0,
      lower quartile=0,
      median=0,
      average = 0.11,
      upper quartile=0.6,
      upper whisker=1.85
    }, color = black!60
    ]coordinates{};
        \addplot+[mark = *, mark options = {black!30},
    boxplot prepared={
      lower whisker=0,
      lower quartile=0.1,
      median=1.07,
      average = 1.34,
      upper quartile=2.12,
      upper whisker=2.77
    }, color = black!60
    ]coordinates{};
        \addplot+[mark = *, mark options = {black!30},
    boxplot prepared={
      lower whisker=0,
      lower quartile=0,
      median=0,
      average = 0.66,
      upper quartile=1.65,
      upper whisker=3.22
    }, color = black!60
    ]coordinates{};
    \end{axis}
\end{tikzpicture}
        \end{center}
    \caption{Boxplots depicting the logarithm of the number of  blocks through their quartiles for all collections. The median (resp. mean) is represented by the band (resp. diamond) inside the box. The ends of the whiskers represent the minimum and the maximum of the values.}
    \label{fig:boxplots}
        \end{figure}

Table \ref{tab:datasets} presents some detailed statistics about each collection. Among these, we report the average number of positive (click, like) feedback and the average number of negative feedback. As we see from the table, datasets {\Out}, {\kasandr}, and {\PANDOR} are the most unbalanced ones in regards to the number of preferred and non-preferred items. With this respect, we also analyzed the distributions of the number of blocks and their size for different collections. Figure \ref{fig:boxplots} shows boxplots representing the logarithm of the number of blocks through their quartiles for all collections. From these plots, it comes out that the distribution of the number of blocks on {\PANDOR}, {\NetF} and {\kasandr} are heavy-tailed with more than the half of the users interacting no more than twice with the system. 

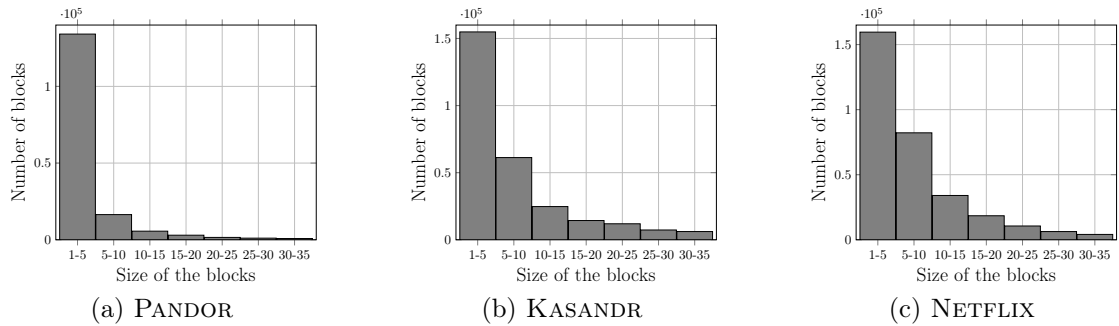
\begin{figure*}[t!]

    \centering
    \begin{tabular}{ccccc}

\begin{tikzpicture}[scale=0.5]
\begin{axis}
[ 
ybar,
 xmajorgrids, 
 bar width = {2.675em},
 yminorticks=true, 
 ymajorgrids, 
 yminorgrids,
 ylabel={Number of blocks},
 xlabel={Size of the blocks},
 ymin = 0.0,
 ymax= 140000.0,
 symbolic x coords={1-5, 5-10, 10-15, 15-20, 20-25, 25-30, 30-35},
 xtick=data,
label style={font=\Large} ,
 ];
 \addplot [fill=gray] coordinates {
      (1-5, 134109)
      (5-10, 16404) 
      (10-15, 5603)
      (15-20, 2975)
      (20-25, 1529)
      (25-30, 1050)
      (30-35, 800)};
\end{axis}
\end{tikzpicture}
&~~~~&
\begin{tikzpicture}[scale=0.5]
\begin{axis}
[ 
    ybar,
 xmajorgrids, 
 bar width = {2.675em},
 yminorticks=true, 
 ymajorgrids, 
 yminorgrids,
 ylabel={Number of blocks},
 xlabel={Size of the blocks},
 ymin = 0.0,
 ymax= 160000.0,
 symbolic x coords={1-5, 5-10, 10-15, 15-20, 20-25, 25-30, 30-35, 35-40, 40-45},
 xtick=data,
label style={font=\Large},
 ];
 \addplot [fill=gray] coordinates {
      (1-5, 154873)
      (5-10, 61243) 
      (10-15, 24733)
      (15-20, 14330)
      (20-25, 11907)
      (25-30, 7281)
      (30-35, 6120)
    };
\end{axis}
\end{tikzpicture}
&~~~~&
\begin{tikzpicture}[scale=0.5]
\begin{axis}[ 
ybar,
 xmajorgrids, 
 bar width = {2.675em},
 yminorticks=true, 
 ymajorgrids, 
 yminorgrids,
 ylabel={Number of blocks},
 xlabel={Size of the blocks},
 ymin = 0.0,
 ymax= 165000.0,
 symbolic x coords={1-5, 5-10, 10-15, 15-20, 20-25, 25-30, 30-35},
 xtick=data,
label style={font=\Large} ,
 ];
 
 \addplot [fill=gray] coordinates {
      (1-5, 159548)
      (5-10, 82173) 
      (10-15, 34034)
      (15-20, 18442)
      (20-25, 10583)
      (25-30, 6300)
      (30-35, 4100)};
 
\end{axis}
\end{tikzpicture}
\\
(a) \PANDOR &~& (b) \kasandr &~& (c) \NetF
\end{tabular}
\caption{Distributions of negative feedback over the train blocks for {\MovieL}, {\PANDOR} and {\kasandr}}
    \label{fig:neg_distribution}

\end{figure*}

Furthermore, we note that on {\PANDOR} the average number of blocks is much smaller than on the two other collections; and that on all three collections the maximum numbers of blocks are $10$0 times more than the average. These plots suggest that a very small number of users (perhaps bots) have an abnormal interaction with the system generating a huge amount of blocks on these three collections. To have a better understanding, Figure \ref{fig:neg_distribution} depicts the number of blocks concerning their size on these three collections. It turns out that on \PANDOR{} the number of blocks having more than $5$ items drops drastically while this number decreases more slowly on the two other collections. As the {\SO} does not sample positive and negative items for updating the weights, it is expected that these updates be performed more often on \PANDOR{} than on the other collections. 

To construct the training and the test sets, we discarded users who did not interact over the shown items and sorted all interactions according to time-based on the existing time-stamps related to each dataset. Furthermore, we considered $80\%$ of each user's first interactions (both positive and negative) for training, and the remaining for the test. Table \ref{tab:detail_setting} resumes the size of the training and the test sets, as well as the percentage of positive items in these sets for all collections. 

\begin{table}[b!]
    \centering
    \begin{tabular}{lllll}
    \hline
    Dataset&$|S_{train}|$&$|S_{test}|$&$pos_{train}$&$pos_{test}$\\
    \hline
    {\ML}-1M&797,758&202,451&58.82&52.39\\
     \Out&1,261,373&336,053&17.64&24.73\\
    \PANDOR&1,579,716&493,663&11.04&12.33\\
    \NetF&3,314,621&873,477&56.27&56.70\\
    \kasandr&12,509,509&3,335,208&3.36&8.56\\
    \hline
    \end{tabular}
    \caption{Number of interactions used for train and test on each dataset, and the percentage of positive feedback among these interactions.}
    \label{tab:detail_setting}
\end{table}

Table \ref{tab:detail_setting} presents the size of the training and the test sets as well as the percentage of positive feedback (preferred items) for all collections ordered by increasing training size. The percentage of positive feedback is inversely proportional to the size of the training sets, attaining $3\%$ for the largest, \kasandr{} collection.

\paragraph{Compared approaches. } To validate the sequential learning approach described in the previous sections, we compared the proposed \SO{} algorithm\footnote{For research purpose, we will publicly make available the code of \SO. } with the following approaches.  
\begin{itemize}
    \item {\MostPop} is a non-learning based approach which consists in recommending the same set of popular items to all users.
    \item Matrix Factorization ({\MF}) \cite{Koren08}, is a factor model which decomposes the matrix of user-item interactions into a set of low dimensional vectors in the same latent space, by minimizing a regularized least square error between the actual value of the scores and the dot product over the user and item representations. The recommendation is then treated as a matrix completion problem by taking the dot product of the user and item
latent factors to fill the missing values.
    \item {\BPR} \cite{rendle_09} corresponds to the model described in the problem statement above (Section \ref{sec:LO}), and {\batch} the batch version of the model which consists in finding  the model parameters $\w=(\vecU,\vecI)$ by minimizing the global ranking loss (Eq.~\ref{eq:RL}). 
   \item {\ProdVec} \cite{GrbovicRDBSBS15},  learns the representation of items using a Neural Networks based model, called word2vec \cite{word_emb}, and performs next-items recommendation using the similarity between the representations of items.
 
    \item {\GRU} \cite{hidasi}  applies recurrent neural network with a GRU architecture for session-based recommendation. The approach also considers the sequence of clicks of the user that depends on all the previous one for learning the model parameters by optimizing a regularized approximation of the relative rank of the relevant item which favors the clicked (preferred) items to be ranked at the top of the list.
\end{itemize}

Hyper-parameters of different models and the dimension of the embedded space for the representation of users and items; as well as the regularisation parameter over the norms of the embeddings for {\SO}, {\BPR}, {\batch} and {\MF} approaches were found by cross-validation.  We fixed $b$ and $B$, used in {\SO}, to respectively the minimum and the average number of blocks found on the training set of each corresponding collection. With the average number of blocks being greater than the median on all collections, the motivation here is to consider the maximum number of blocks by preserving the model from the bias brought by the too many interactions of the very few number of users.

\begin{table}[!b]
\centering
{\begin{tabular}{c|ccc|ccc}
\hline

\cline{2-7}
\multirow{3}{*}{Dataset}&\multicolumn{6}{c}{Test Loss, Eq.~\eqref{eq:RL}}\\
\cline{2-7}
&\multicolumn{3}{c|}{30 min}&\multicolumn{3}{c}{1 hour}\\
\cline{2-7}
&\batch&{\BPR}&\SO&\batch&{\BPR}&\SO\\
\hline
\ML-1M &0.751&0.678&\bf{0.623}&0.744&0.645&\bf{0.608}\\
\Out &0.753&0.650&\bf{0.646}& 0.747&0.638&\bf{0.635}\\
\PANDOR &0.715&0.671&\bf{0.658}&0.694&0.661&\bf{0.651}\\
\NetF &0.713&0.668&\bf{0.622}&0.694&0.651&\bf{0.614}\\
\kasandr &0.663&0.444&\bf{0.224}&0.631&0.393&\bf{0.212}\\
\hline
\end{tabular}
}
\caption{Comparison between \BPR, \batch{} and \SO{} approaches in terms on test loss after $30$ minutes and $1$ hour of training.}
\label{test_loss}
\end{table}

\paragraph{Evaluation setting and results. }

We begin our comparisons by testing {\batch}, \BPR{} and \SO{} approaches over the logistic ranking loss (Eq. \ref{eq:instloss}) which is used to train them. Results on the test, after training the models 30 minutes and 1 hour are shown in Table \ref{test_loss} and best performance is in bold. {\batch} (resp. {\SO}) techniques have the worse (resp. best) test loss on all collections, and the difference between their performance is larger for bigger size datasets. 
These results suggest that the local ranking between preferred and no-preferred items present in the blocks of the training set better reflects the preference of users than the ranking of random pairs of items or their global ranking without this contextual information. Furthermore, as in {\SO} updates occur after the creation of a block, and that most of the blocks contain very few items (Figure \ref{fig:neg_distribution}),  weights are updated more often than in {\BPR} or {\batch}. This is depicted in Figure \ref{fig:losses} which shows the evolution of the training error over time for {\batch}, \BPR{} and \SO{} on all collections. As we can see, the training error decreases in all cases, and theoretically, the three approaches converge to the same minimizer of the ranking loss (Eq. \ref{eq:RL}). However, the speed of convergence is much faster with {\SO}.

\begin{flushleft}
\begin{figure}[b!]
\small
    \centering
    \begin{tabular}{cc}

\begin{tikzpicture}[scale=0.45]
\begin{axis}[ 
 width=1.0\columnwidth, 
 height=0.65\columnwidth, 
 xmajorgrids, 
 yminorticks=true, 
 ymajorgrids, 
 yminorgrids,
 ylabel={Training error ~$\mathcal{L}(\omega)$},
 xlabel = {Time, min.},
 ymin = 0.6,
 ymax= 0.77,
 xmin = 0,
 xmax = 60,
label style={font=\Large} ,
 ];

 \addplot  [color=black,
                dash pattern=on 1pt off 3pt on 3pt off 3pt,
                mark=none,
                mark options={solid},
                smooth,
                line width=1.2pt]  file { ./ml_batch.txt}; 
 \addlegendentry{ \batch };

  \addplot  [color=black,
                dashed,
                mark=none,
                mark options={solid},
                smooth,
                line width=1.2pt]  file { ./ml_sgd.txt }; 
 \addlegendentry{ \BPR };
 
  \addplot  [color=black,
                dotted,
                mark=none,
                mark options={solid},
                smooth,
                line width=1.2pt]  file { ./ml_online.txt }; 
 \addlegendentry{ \SO };

\end{axis}
\end{tikzpicture}
&
\begin{tikzpicture}[scale=0.45]
\begin{axis}[ 
 width=\columnwidth, 
 height=0.65\columnwidth, 
 xmajorgrids, 
 yminorticks=true, 
 ymajorgrids, 
 yminorgrids,
 ylabel={Training error ~$\mathcal{L}(\omega)$},
 xlabel = {Time, min},
 ymin = 0.60,
 ymax=0.77,
 xmin = 0,
 xmax = 60,
label style={font=\Large} ,
tick label style={font=\Large}
 ];

  \addplot  [color=black,
                dash pattern=on 1pt off 3pt on 3pt off 3pt,
                mark=none,
                mark options={solid},
                smooth,
                line width=1.2pt]  file { ./out_batch.txt }; 
 \addlegendentry{ \batch };

  \addplot  [color=black,
                dashed,
                mark=none,
                mark options={solid},
                smooth,
                line width=1.2pt]  file { ./out_sgd.txt }; 
 \addlegendentry{ \BPR };
  \addplot  [color=black,
                dotted,
                mark=none,
                mark options={solid},
                smooth,
                line width=1.2pt]  file { ./out_online.txt }; 
 \addlegendentry{ \SO };
\end{axis}
\end{tikzpicture}\\
(a) \ML-1M & (b) \Out\\
\begin{tikzpicture}[scale=0.45]
\begin{axis}[ 
 width=1.0\columnwidth, 
 height=0.65\columnwidth, 
 xmajorgrids, 
 yminorticks=true, 
 ymajorgrids, 
 yminorgrids,
 ylabel={Training error ~$\mathcal{L}(\omega)$},
 xlabel = {Time, min.},
 ymin = 0.55,
 ymax= 0.78,
 xmin = 0,
 xmax = 60,
label style={font=\Large} ,
tick label style={font=\Large}
 ]
 
 \addplot  [color=black,
                dash pattern=on 1pt off 3pt on 3pt off 3pt,
                mark=none,
                mark options={solid},
                smooth,
                line width=1.2pt]  file { ./pandor_batch.txt }; 
 \addlegendentry{ \batch }; 
 
   \addplot  [color=black,
                dashed,
                mark=none,
                mark options={solid},
                smooth,
                line width=1.2pt]  file { ./pandor_sgd.txt }; 
 \addlegendentry{ \BPR };

  \addplot  [color=black,
                dotted,
                mark=none,
                mark options={solid},
                smooth,
                line width=1.2pt]  file { ./pandor_online.txt }; 
 \addlegendentry{ \SO };
\end{axis}
\end{tikzpicture}
&
\begin{tikzpicture}[scale=0.45]
\begin{axis}[ 
 width=\columnwidth, 
 height=0.65\columnwidth, 
 xmajorgrids, 
 yminorticks=true, 
 ymajorgrids, 
 yminorgrids,
 ylabel={Training error ~~~~$\mathcal{L}(\omega)$},
 xlabel = {Time, min},
 ymin = 0.6,
 ymax=0.76,
 xmin = 0,
 xmax = 60,
label style={font=\Large} ,
tick label style={font=\Large}
 ];
 
  \addplot  [color=black,
                dash pattern=on 1pt off 3pt on 3pt off 3pt,
                mark=none,
                mark options={solid},
                smooth,
                line width=1.2pt]  file { ./netflix_batch.txt }; 
 \addlegendentry{ \batch };

  \addplot  [color=black,
                dashed,
                mark=none,
                mark options={solid},
                smooth,
                line width=1.2pt]  file { ./netflix_sgd.txt }; 
 \addlegendentry{ \BPR };
  \addplot  [color=black,
                dotted,
                mark=none,
                mark options={solid},
                smooth,
                line width=1.2pt]  file { ./netflix_online.txt }; 
 \addlegendentry{ \SO };
 \end{axis}
\end{tikzpicture}
\\
(c) \PANDOR & (d) \NetF\\
\multicolumn{2}{c}{

\begin{tikzpicture}[scale=0.45]
\begin{axis}[ 
 width=\columnwidth, 
 height=0.65\columnwidth, 
 xmajorgrids, 
 yminorticks=true, 
 ymajorgrids, 
 yminorgrids,
 ylabel={Training error ~$\mathcal{L}(\omega)$},
 xlabel = {Time, min},
 ymin = 0.4,
 ymax=0.78,
 xmin = 0,
 xmax = 60,
label style={font=\Large} ,
tick label style={font=\Large}
 ];

 \addplot  [color=black,
                dash pattern=on 1pt off 3pt on 3pt off 3pt,
                mark=none,
                mark options={solid},
                smooth,
                line width=1.2pt]  file { ./kassandr_batch.txt }; 
 \addlegendentry{ \batch };

  \addplot  [color=black,
                dashed,
                mark=none,
                mark options={solid},
                smooth,
                line width=1.2pt]  file { ./kassandr_sgd.txt }; 
 \addlegendentry{ \BPR };
 
  \addplot  [color=black,
                dotted,
                mark=none,
                mark options={solid},
                smooth,
                line width=1.2pt]  file { ./kassandr_online.txt }; 
 \addlegendentry{ \SO };

\end{axis}
\end{tikzpicture}
}\\
\multicolumn{2}{c}{(e) \kasandr}
\end{tabular}
\caption{Evolution of the loss on training sets for both {\batch}, {\BPR} and {\SO} as a function of time in minutes for all collections.}
    \label{fig:losses}
\end{figure}
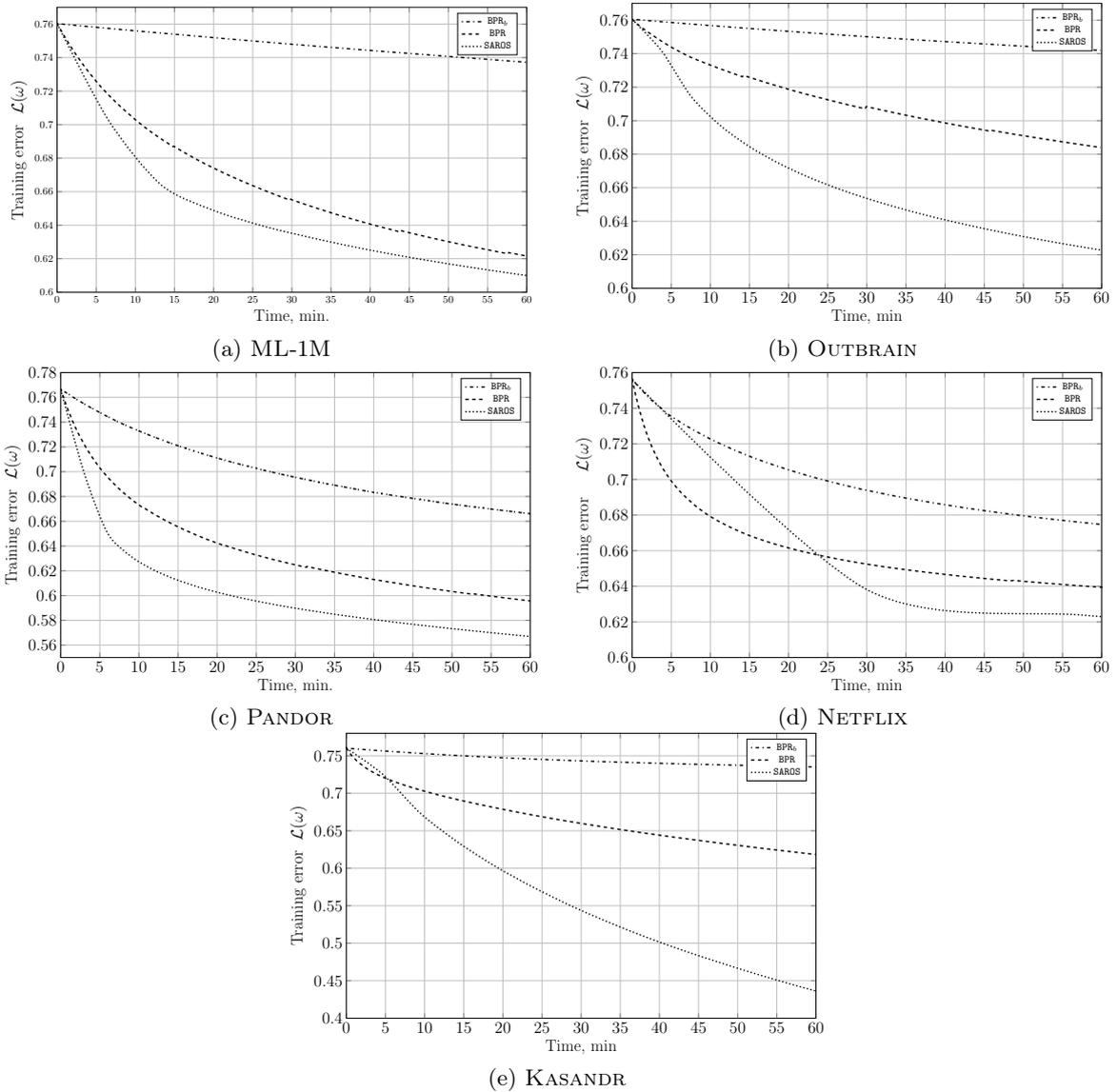
\end{flushleft}
\vspace{-8mm}We also compare the performance of all the approaches on the basis of the common ranking metrics, which are the Mean Average Precision at rank $K$ (\mapk) over all users  defined as $\mapk=\frac{1}{N}\sum_{u=1}^{N}\apk(u)$, where $\apk(u)$ is the average precision of  preferred items of user $u$ in the top $K$ ranked ones; and the Normalized Discounted Cumulative Gain at rank $K$ (\ndcgk) that computes the ratio of the obtained ranking to the ideal case and allow to consider not only binary relevance as in Mean Average Precision, $\ndcgk = \frac{1}{N}\sum_{u=1}^{N}\frac{\dcgk(u)}{\idcgk(u)}$, where $\dcgk(u) = \sum_{i=1}^{K}\frac{2^{rel_{i}}-1}{\log_{2}(1+i)}$, $rel_{i}$ is the graded relevance of the item at position $i$; and $\idcgk(u)$ is $\dcgk(u)$ with an ideal ordering equals to $\sum_{i=1}^{K}\frac{1}{\log_{2}(1+i)}$ for $rel_{i}\in[0,1]$  \cite{Manning:2008}.

\begin{table*}[t!]
    \centering
     \resizebox{\textwidth}{!}{\begin{tabular}{c|ccccccc|ccccccc}
    \hline
     \cline{2-15}
     \multirow{2}{*}{Dataset}&\multicolumn{7}{c}{\mapfive}&\multicolumn{7}{c}{\mapten}\\
     \cline{2-15}
     &{\MostPop}&\ProdVec&\MF&{\batch}&{\BPR}&{\GRU}&{\SO}&{\MostPop}&\ProdVec&\MF&{\batch}&{\BPR}&{\GRU}&{\SO}\\
     \hline
     \ML-1M &.074&.793&.733&.713&\emph{.836}&.777&\bf{.837}&.083&.772&.718&.688&\emph{.807}&.750&\bf{.808}\\
     \Out&.007&.228&.531&.477&\emph{.573}&.513&\bf{.619}&.009&.228&.522&.477&.\emph{563}&.509&\bf{.607}\\
     \PANDOR&.003&.063&.266&.685&\emph{.744}&.673&\bf{.750}&.004&.063&.267&.690&\emph{.746}&.677&\bf{.753}\\
    \NetF&.039&.699&.793&.764&.865&.774&\bf{.866}&.051&.690&.778&.748&.845&.757&\bf{.846}\\
     \kasandr&.002&.012&.170&.473&.507&\emph{.719}&\bf{.732}&.3e-5&.012&.176&.488&.521&\emph{.720}&\bf{.747}\\
     \hline
    \end{tabular}
    }
    \caption{Comparison between \MostPop, \ProdVec, \MF, \batch, \BPR{} and \SO{} approaches in terms on \mapfive and \mapten. }
    \label{tab:online_vs_all_1h}
\end{table*}

\begin{table*}[t!]
    \centering
     \resizebox{\textwidth}{!}{\begin{tabular}{c|ccccccc|ccccccc}
    \hline
     \cline{2-15}
     \multirow{2}{*}{Dataset}&\multicolumn{7}{c}{\ndcgfive}&\multicolumn{7}{c}{\ndcgten}\\
     \cline{2-15}
     &{\MostPop}&\ProdVec&\MF&{\batch}&{\BPR}&{\GRU}&{\SO}&{\MostPop}&\ProdVec&\MF&{\batch}&{\BPR}&{\GRU}&{\SO}\\
     \hline
     \ML-1M &.090&.758&.684&.652&\emph{.786}&.721&\bf{.788}&.130&.842&.805&.784&\emph{.873}&.833&\bf{.874}\\
     \Out &.011&.232&.612&.583&.\emph{671}&.633&\bf{.710}&.014&.232&.684&.658&\emph{.724}&.680&\bf{.755}\\
     \PANDOR &.005&.078&.300&.874&\emph{.899}&.843&\bf{.903}&.008&.080&.303&.890&\emph{.915}&.862&\bf{.913}\\
    \NetF &.056&.712&.795&.770&.864&.777&\bf{.865}&.096&.770&.834&.849&.913&.854&\bf{.914}\\
     \kasandr &.002&.012&.197&.567&.603&\emph{.760}&\bf{.791}&.002&.012&.219&.616&.650&\emph{.782}&\bf{.815}\\
     \hline
    \end{tabular}
    }
    \caption{Comparison between \MostPop, \ProdVec, \MF, \batch, \BPR{} and \SO{} approaches in terms on \ndcgfive and \ndcgten. }
    \label{tab:online_vs_all_ndcg_1h}
\end{table*}

Table \ref{tab:online_vs_all_1h} (resp. Table \ref{tab:online_vs_all_ndcg_1h}) presents {\mapfive} and {\mapten} (resp. {\ndcgfive} and {\ndcgten})  performance measures of  {\MostPop}, \ProdVec, \MF, {\GRU}, {\batch}, {\BPR} and {\SO} after 1 hour of training, over the test sets of the different datasets. The non-machine learning method, {\MostPop},  gives results of an order of magnitude lower than the learning based approaches. Moreover,  the factor model {\MF} which predicts clicks by matrix completion is less effective when dealing with implicit feedback than ranking based models especially on large datasets where there are fewer interactions. We also found that embeddings found by ranking based models, in the way that the user preference over the pairs of items is preserved in the embedded space by the dot product, are more robust than the ones found by {\ProdVec}. Also, it comes out that {\SO} is the most competitive approach, performing better -- and in some cases even outperforming -- other approaches over all collections. When comparing {\GRU} with {\BPR} and {\batch} that also minimize the same surrogate ranking loss, the former outperforms both approaches. This is mainly because  {\GRU} optimizes an approximation of the relative rank that favors interacted items to be in the top of the ranked list while the logistic ranking loss, which is mostly related to the Area under the ROC curve \cite{Usunier:1121}, pushes up clicked items for having good ranks in average. However, the minimization of the logistic ranking loss over blocks of very small size pushes the clicked item to be ranked higher than the no-clicked ones in several lists of small size and it has the effect of favoring the clicked item to be at the top of the whole merged lists of items. Furthermore, by discarding users who do not have the same behavior regarding interaction than the majority of users, {\SO} performs better than {\GRU}. The reason might be that the approaches which minimize a ranking measure that favors relevant elements to be at the top of the list are much more sensible to noisy data than those which optimize an approximation of the AUC. By discarding users who are suspected to add noise concerning the interactions of the majority of users, {\SO} becomes highly competitive with respect to {\GRU}. 

\section{Conclusion}\label{sec:Conclusion}

The contributions of this paper are twofold. First, we proposed {\SO},  a novel learning framework for large-scale Recommender Systems that sequentially updates the weights of a ranking function user by user over blocks of items ordered by time where each block is a sequence of negative items followed by a last positive one. The main hypothesis of the approach is that the preferred and no-preferred items within a local sequence of user interactions express better the user preference than when considering the whole set of preferred and no-preferred items independently one from another. The approach updates the model parameters user per user over blocks of items constituted by a sequence of unclicked items followed by a clicked one. The parameter updates are discarded for users who interact very little or a lot with the system.  The second contribution is a theoretical analysis of the proposed approach which bounds the deviation of the ranking loss concerning the sequence of weights found by the algorithm and its minimum in the case where the loss is convex.  Empirical results conducted on five real-life implicit feedback datasets support our founding and show that the proposed approach is significantly faster than the common batch and online optimization strategies that consist in updating the parameters over the whole set of users at each epoch, or after sampling random pairs of preferred and no-preferred items. The approach is also shown to be highly competitive concerning state of the art approaches on {\mapk} and {\ndcgk}.

\section*{Acknowledgements}

A major part of this work was done while YM visited EPFL, Bernoulli center, and Grenoble Informatics Laboratory, AMA team. The work at LANL was carried out under the auspices of the National Nuclear Security Administration of the U.S. Department of Energy under Contract No. DE-AC52-06NA25396. The work was partially supported by the Laboratory Directed Research and Development program of LANL under project number 20190351ER, also DOE/OE/GMLC and LANL/LDRD/CNLS projects.

\bibliography{icml-silicom}
\bibliographystyle{chicago}

\appendix

\section{Proofs}

We summarize the notations used in the Supplementary part in Table~\ref{table:notation}. We omit ${\cal D}$, ${\cal D}_u$, ${\cal D}_b^{\cal B}$, and ${\cal D}_{{\cal B}, u}$ in the notation of an expectation and a variance if the underlying distribution is clear from the context. 

\begin{table}[h!]
\centering
 \begin{tabular}{l|l} 
 ${\cal D}$ & joint distribution over users and items\\ 
 \hline
 ${\cal D}_{u}$ & conditional distribution of items for \\
 & a fixed user $u$\\\hline
 ${\cal D}_{{\cal B}_u}$ & conditional distribution of items for\\
 & a fixed user $u$ and block ${\cal B}$\\
 \hline
 ${\cal D}^{\cal B}_{u}$ & conditional distribution of blocks of \\
 & positive/negative items for a fixed user $u$\\
 \hline 
 $\posI_u$ & positive feedbacks for user $u$\\
 \hline
 $\negI_u$ & negative feedbacks for user $u$\\
 \hline 
 ${\cal B}_u^l$ & $l$-th block considered for user $u$\\
 \hline
 $\Pi_u^{\ell}$ & number of positive feedbacks for user $u$\\
 \hline
 $\text{N}_u^{\ell}$ & number of negative feedbacks for user $u$\\
 \hline
 $\ell_{u, i, i'}(\omega)$ & Loss over user $u$ and a pait of items $(i, i')$\\
 \hline 
 $\widehat{\Loss}_u(\w)$ & Empirical loss with respect to user $u$\\
 & $\widehat{\Loss}_u(\w) = \frac{1}{|\posI_u||\negI_u|}\!\sum_{i\in \posI_u}\!\sum_{i'\in \negI_u} \!\ell_{u,i,i'} (\w)$\\
 \hline 
 ${\widehat {\cal L}}_{\bloc^\ell_u}(\w)$ &  Empirical loss with respect to a block of items \\ 
 & ${\widehat {\cal L}}_{\bloc^\ell_u}(\w) = \frac{1}{|\Pi_u^{\ell}||\text{N}_u^{\ell}|}\sum_{i \in \Pi_u^{\ell}} \sum_{i'\in \text{N}_u^{\ell}} \ell_{u, i, i'} (\w)$ \\
 \hline
 ${\cal L}(\omega)$ & Expected loss of the classifier, ${\cal L}(\omega) = \mathbb{E}_{{\cal D}_u}\widehat{\cal L}_u(\omega)$
 \end{tabular}
 \caption{Notation used in the proofs.}
 \label{table:notation}
\end{table}

\setcounter{theorem}{0}

For the sake of self-consistency, recall the assumption we use as a starting point of the proofs. 

Prior to a proof of the theorem, we propose a technical lemma
\begin{lemma}\label{lem:tech}
Let a sequence of items $(i_1, \dots, i_m)$ generated generated i.i.d.  according to a distribution~${\cal D}_{u}$ over items for a given user $u$. Then for any sequence of blocks $\{{\cal B}^1_u, \dots, {\cal B}^k_u\}$ generated by~{\SO} algorithm for that user:
\begin{gather}\label{lem1:eq1}
\mathbb{E}_{{\cal D}_u} \left[\frac{1}{k}\sum_{l=1}^k\nabla {\widehat {\cal L}}_{{\cal B}^l_u}(\w)\right] = \nabla \widehat{\Loss}_u(\w), 
\quad \text{ with } \widehat{\Loss}_u(\w)\!=\!\frac{1}{|\posI_u||\negI_u|}\!\sum_{i\in \posI_u}\!\sum_{i'\in \negI_u} \!\ell_{u,i,i'} (\w),
\end{gather}
where 
\[
{\widehat {\cal L}}_{\bloc^\ell_u}(\w) = \frac{1}{|\Pi_u^{\ell}||\text{N}_u^{\ell}|}\sum_{i \in \Pi_u^{\ell}} \sum_{i'\in \text{N}_u^{\ell}} \ell_{u, i, i'} (\w), \]
and $\Pi_u^l$, $N_u^l$ are the sets of positive (resp. negative) interactions in the block.

In other words, the  expected gradient of empirical loss, taken over random blocks ${\cal B}_1, 
\dots, {\cal B}_k$ generated by the ${\SO}$ algorithm for a user $u$, equals to the expected loss over $u$. Moreover, if for any $(u, i, i')$ one has $\|\nabla \ell_{u, i, i'}(\omega) \| \le \gamma^2$, then 
\begin{gather*}
\mathbb{E}_{{\cal D}_u} \biggl\|\nabla \widehat{\cal L}_u(\omega) - \frac{1}{k}\sum_{\ell=1}^k \nabla {\widehat {\cal L}}_{\bloc_u^\ell}(\w) \biggr\|_2^2 \le 3\frac{\gamma^2}{k}. 
\end{gather*}
\end{lemma}
%
%
\begin{Proof} 
Consider the expectation of the gradient of the empirical loss over a user $u$, $\nabla {\cal L}_{{\cal B}^l_u}(\omega)$, taken with respect to a block ${\cal B}^l$. For a fixed block, ${\cal B}^l$, the value of $|N_u|\cdot|\Pi_u|$ is a constant. Thus, due to the linearity of expectation, for the sum of random $\ell_{u, i, i'} (\w)$ we have
\begin{gather}\label{eq:exp-lin}
    \mathbb{E}_{{\cal D}_{{\cal B}_u^l}}\!\!\nabla \widehat{\cal L}_{{\cal B}^l_u}(\omega) = \mathbb{E}_{{\cal D}_{{\cal B}_u^l}} \!\!\!\left[ \frac{1}{|\Pi_u^{\ell}||\text{N}_u^{\ell}|}\sum_{i \in \Pi_u^{\ell}}\! \sum_{i'\in \text{N}_u^{\ell}} \!\!\nabla\ell_{u, i, i'} (\w) \right] \!\! = \!\! \frac{1}{|\Pi_u^{\ell}||\text{N}_u^{\ell}|}\!\!\sum_{i \in \Pi_u^{\ell}} \!\sum_{i'\in \text{N}_u^{\ell}} \!\!\nabla \widehat{\cal L}_u(\omega) = \nabla \widehat{\cal L}_u(\omega)
\end{gather}
where the first sum consists of a non-zero number of addends as 
each block contains at least one positive and one negative item. 

Thus, by the law of total expectation, $\mathbb{E}_{\psi} f(\psi) = \mathbb{E}_\eta \mathbb{E}_{\psi|\eta} f(\psi)$ for any properly defined random variables $\psi$, $\eta$ and a function $f$,  we have 
\begin{align*}
  \mathbb{E}_{{\cal D}_u} 
    \left[
        \frac{1}{k}\sum_{l=1}^k \nabla {\widehat {\cal L}}_{\bloc^l}(\w)
    \right] 
    & = 
    \frac{1}{k} \mathbb{E}_{{\cal D}_u} 
    \left[
        \sum_{l=1}^k \nabla {\widehat {\cal L}}_{\bloc^l}(\w)
    \right]
    \\
    & = 
    \frac{1}{k}\sum_{l=1}^k \mathbb{E}_{{\cal D}_u^{{\cal B}^l_u}}\mathbb{E}_{{\cal D}_{{\cal B}^l_u}} 
    \left[
         \nabla {\widehat {\cal L}}_{{\bloc^l_u}}(\w)
    \bigg| \bloc^l_u\right] 
     = 
    \frac{1}{k}\sum_{l=1}^k \mathbb{E}_{{\cal D}_u} \nabla \widehat{\cal L}_u(\omega) = \nabla \widehat{\cal L}_u(\omega)
\end{align*}    
where the last is due to Eq.~\eqref{eq:exp-lin}. 

To proof the bound on variance, recall, that {\SO} constructs the blocks sequentially, so that the number of  positive and negative items in any block ${\cal B}$ is affected only by the previous and the next block. Thus, any block after the next to ${\cal B}$ and before the previous to ${\cal B}$ are conditionally independent for any fixed ${\cal B}$. Then if $V^2 = \mathbb{E}_{{\cal D}_{\cal B}} \|\nabla \widehat{\cal L}_{{\cal B}_l} (\omega) - \nabla \widehat{\cal L}_u (\omega)\|_2^2$ one has: 
\begin{align*}
    \mathbb{E}_{{\cal D}^{{\cal B}^1_u}, \dots, \dots, {\cal D}^{{\cal B}^k_u}} \hspace{-7mm}
    &
    \hspace{7mm}\biggl\|\frac{1}{k}\sum_{j=1}^k \left(\nabla \widehat{\cal L}_{{\cal B}^j_u} (\omega) - \nabla \widehat{\cal L}_u(\omega)\right)\biggr\|_2^2 \\
    & = \mathbb{E}_{{\cal D}^{{\cal B}^1_u}, \dots, \dots, {\cal D}^{{\cal B}^k_u}} \left[\frac{1}{k^2} \sum_{i,j=1}^k \biggl(\nabla\widehat{\cal L}_{{\cal B}^i_u} (\omega) - \nabla \widehat{\cal L}_u(\omega)\biggr)\biggl(\nabla \widehat{\cal L}_{{\cal B}^j_u} (\omega) - \nabla \widehat{\cal L}_u(\omega)\biggr)^\top\right]\\
    & = \mathbb{E}_{{\cal D}_{{\cal B}^2_u}}\mathbb{E}_{{{\cal D}_{{\cal B}^1_u}}, {{\cal D}_{{\cal B}^3_u}} \dots, {{\cal D}_{{\cal B}^k_u}}| {\cal B}_2} \left[\frac{1}{k^2} \sum_{i,j=1}^k \biggl(\nabla\widehat{\cal L}_{{\cal B}^i_u} (\omega) - \nabla \widehat{\cal L}_u(\omega)\biggr)\biggl(\nabla \widehat{\cal L}_{{\cal B}^j_u} (\omega) - \nabla \widehat{\cal L}_u(\omega)\biggr)^\top\bigg| {\cal B}_2\right] \\
    & \le \frac{3V^2}{k^2} + \frac{1}{k^2}\mathbb{E}_{{{\cal D}^{{\cal B}^1_u}, {\cal D}^{{\cal B}^3_u}, \dots, {\cal D}^{{\cal B}^k_u}}}
    \sum_{\substack{i,j=1\\ i, j\neq 2}}^k \biggl(\nabla\widehat{\cal L}_{{\cal B}^i_u} (\omega) - \nabla \widehat{\cal L}_u(\omega)\biggr)\biggl(\nabla \widehat{\cal L}_{{\cal B}^j_u} (\omega) - \nabla \widehat{\cal L}_u(\omega)\biggr)^\top \le \frac{3V^2}{k}
\end{align*}
To conclude the proof it remains to note that $V^2 \le \gamma^2$ as 
$V^2 \le \mathbb{E}_{\cal B} \|\nabla \widehat{\cal L}_{{\cal B}_u}\|_2^2$. 
\end{Proof}

\begin{theorem}
Let $\ell_{u, i, i'}(\omega)$ and ${\cal L}(\omega)$ satisfy Assumption~\ref{asmp:smooth}.
Then for any constant step size $\eta$, verifying $0< \eta \le \min\{1/(\beta B), 1/\sqrt{NB(\sigma^2 + 3\gamma^2/b)}\}$, and any set of users $\userS\doteq [N]$; algorithm {\SO} iteratively generates  a sequence $\{\omega_j^0\}_{u\in \userS}$ such that 
\begin{gather*}
    \frac{1}{\beta} \mathbb{E}_{\cal D}\|\nabla {\cal L}({\omega}_u^0)\|_2^2
    \le
    \frac{\beta B \Delta_{\cal L}^2}{u} + 2\Delta_{\cal L}\sqrt{\frac{B\sigma^2 + 3B\gamma^2/b}{u}}, \quad \Delta_{\cal L}^2 = {\frac{2}{\beta}({\cal L}(\omega_0) - {\cal L}(\omega^*))}
\end{gather*}
where the expectation is taken with respect to users chosen randomly according to the uniform distribution $p_u=\frac{1}{N}$.

Furthermore, if the ranking loss ${\cal L}(\omega)$ is convex, then for any $\bar\omega_u = \sum_{j\le u} \omega_j^0$ we have
\begin{gather*}
    {\cal L}({\bar \omega}_u) - {\cal L}({\omega_*}) \le \frac{\beta B \Delta_\omega^2}{u} + 2\Delta_{\omega}\sqrt{\frac{B\sigma^2 + 3B\gamma^2/b}{u}}, \quad \Delta_{\omega}^2 = \|\omega_0 - \omega_*\|_2^2.
\end{gather*}
\end{theorem}

Proof of the theorem is mainly based on the randomized stochastic gradient descent analysis~\cite{ghadimi2013stochastic}.

\begin{Proof}
Let $g_u^t$ be a gradient of the loss function taken for user $u$ over block ${\cal B}^t_u$:
\[
    g_u^t = \frac{1}{|N_u^t||\Pi_u^t|} \sum_{i \in N_u^t, i' \in \Pi_u^t} \nabla \ell_{u, i, i'} (\omega_u^{t-1}),
\]
By Lemma~\ref{lem:tech} we have $\mathbb{E}_{{\cal D}_{{\cal B}^t_u}}\,  g_u^t = \nabla \hat {\cal L}_u(\omega)$. In the notation of Algorithm {\SO},
\[
    \omega_u^{t+1} = \omega_u^t - \eta g_u^t,
        \qquad \omega_{u+1}^{0} = \omega_u^{|{\cal B}_u|}, \qquad
    \omega_{u+1}^{0} - \omega_u^{0} = \eta \sum_{t\in {\cal B}_u} g_u^t.
\]
Let $\delta_u^t = g_u^t - \nabla {\cal L}(\omega_u^{0})$, and let ${\cal B}_u$ be a set of all blocks correspond to user $u$. Using the smoothness of the loss function implied by Assumption~\ref{asmp:smooth} one has for ${\omega}_{u+1}^0$:
\begin{align}\label{eq:proof01}
    {\cal L}(\omega_{u+1}^0)
    &
        \le {\cal L}(\omega_{u}^0) - \eta \langle \nabla {\cal L}(\omega_{u}^0), \omega_{u+1}^0 - \omega_{u}^0\rangle + \frac{\beta}{2} \eta^2 \left\|\sum_{t \in {\cal B}_u} g_u^t\right\|_2^2 \noindent\\
    &
        = {\cal L}(\omega_{u}^0) - \eta \sum_{t \in {\cal B}_u}\langle \nabla {\cal L}(\omega_{u}^0), g_u^t \rangle + \frac{\beta}{2} \eta^2 \left\|\sum_{t \in {\cal B}_u} g_u^t\right\|_2^2 \noindent\\
    &
        = {\cal L}(\omega_{u}^0) - \eta |{\cal B}_u| \|\nabla {\cal L}(\omega_{u}^0)\|_2^2 - \eta \sum_{t \in {\cal B}_u}\langle\nabla {\cal L}(\omega_{u}^0), \delta_u^t\rangle \noindent\\
    &   \hspace{18
    mm}+\frac{\beta}{2} \eta^2 \left[|{\cal B}_u|^2\|\nabla {\cal L}(\omega_{u}^0)\|_2^2 + 2|{\cal B}_u|\sum_{t \in {\cal B}_u}\langle\nabla {\cal L}(\omega_{u}^0), \delta_u^t\rangle + \sum_{t\in {\cal B}_u}\|\delta_u^t\|^2\right]\noindent\\
    &
        = {\cal L}(\omega_{u}^0) - \left(\hat\eta_u - \frac{\beta}{2} {\hat\eta}^2_u \right) \|\nabla {\cal L}(\omega_{u}^0)\|_2^2  \noindent\\
    &
        \hspace{32mm} - (\hat\eta_u - \beta {\hat\eta}^2_u) \sum_{t \in {\cal B}_u}\left\langle \nabla {\cal L}(\omega_{u}^0), \frac{\delta_u^t}{|{\cal B}_u|}\right\rangle + \frac{\beta}{2}{\hat\eta}^2_u \sum_{t\in {\cal B}_u} \left\|\frac{\delta_u^t}{|{\cal B}_u|}\right\|_2^2 
\end{align}
where $\hat\eta_u = |{\cal B}_u|\eta$.
\\~\\~\\

Then re-arranging and summing up, we have
\begin{align*}
    \sum_{u=1}^N & \left(\hat\eta_u -  \frac{\beta}{2}{\hat\eta}^2_u\right) \|\nabla {\cal L}(\omega_u)\|_2^2 \\
    & \le {\cal L}(\omega_u) - {\cal L}(\omega^*) - \sum_{u=1}^{N} (\hat\eta_u - \beta\hat\eta^2_u)\left\langle\nabla{\cal L}(\omega_u), \sum_{t\in {\cal B}_u} \frac{\delta_u^t}{|{\cal B}_u|}\right\rangle + \frac{\beta}{2}\sum_{u=1}^N \hat\eta_u^2 \left\|\sum_{t\in {\cal B}_u}\frac{\delta_u^t}{|{\cal B}_u|}\right\|_2^2
\end{align*}

By Lemma~\ref{lem:tech}, the stochastic gradient taken with respect to a block of items gives an unbiased estimate of the gradient, thus
\begin{align}\label{eq:thm01-proof02}
    \mathbb{E}_{{\cal D}_u}\biggl[\biggl\langle\nabla{\cal L}(\omega_u), \sum_{t\in {\cal B}_u} \frac{\delta_u^t}{|{\cal B}_u|}\biggr\rangle \bigg| \xi_u\biggr] = 0,
\end{align}
where $\xi_u$ is a set of users preceding $u$. As in the conditions of the theorem $b\le {\cal B}_u$ almost surely, one has by Lemma~\ref{lem:tech} and the law of total variation, $\Var \psi = \mathbb{E}[\Var(\psi|\eta)] + \Var[\mathbb{E}[\psi|\eta]]$:
\begin{gather}\label{eq:thm01-proof03}
    \mathbb{E}_{{\cal D}_u}\,\left\|\sum_{t\in {\cal B}_u}\frac{\delta_u^t}{|{\cal B}_u|}\right\|_2^2 \le \sigma^2 + \frac{3\gamma^2}{b} 
\end{gather}
where the first attend on the right-hand side of Eq.~\eqref{eq:thm01-proof03} comes from Assumption~\ref{asmp:smooth}, and the second term is due to Lemma~\ref{lem:tech}. 

Finally, one obtains
\begin{gather*}
    \sum_{u=1}^N \left(\hat\eta_u -  \frac{\beta}{2}{\hat\eta}^2_u\right) \mathbb{E}_{\xi_N}\,\|\nabla {\cal L}(\omega_u)\|_2^2 \le {\cal L}(\omega_0) - {\cal L}(\omega^*) + \frac{\beta(\sigma^2b + 3\gamma^2)}{2b} \sum_{u=1}^N \hat\eta_u^2.
\end{gather*}
Condition $\beta \eta B \le 1$ implies $\hat\eta_u - {\beta}{\hat\eta}^2_u/2 \ge \hat\eta_u/2$, thus
\begin{align*}
    \frac{1}{\beta}\mathbb{E}_{\cal D}\,\|\nabla {\cal L}(\omega)\|_2^2 \le
    \frac{1}{\sum_{u=1}^N\hat\eta_u}\left[
    \frac{2({\cal L}(\omega_0) - {\cal L}(\omega_*))}{\beta} + \left(\sigma^2 + 3\frac{\gamma^2}{b}\right) \sum_{u=1}^N\hat\eta^2_u
    \right]
\end{align*}
Taking
\[
    \eta = \min\left\{\eta_1, \psi\eta_2\right\}, \quad \eta_1 = \frac{1}{\beta B}, \quad \eta_2 = \frac{1}{\sqrt{NB(\sigma^2 + 3\gamma^2/b)}}
\]
for some $\psi > 0$. Let $D_{\cal L} = \sqrt{2({\cal L}(\omega_0) - {\cal L}(\omega_*))/\beta}$, then
\begin{align*}
    \frac{1}{\beta}\mathbb{E}_{\cal D}\,\|\nabla {\cal L}(\omega)\|_2^2
    & \le
    \frac{D_{\cal L}^2}{N \min\{\eta_1, \psi\eta_2\}} + \left(\sigma^2 + 3\frac{\gamma^2}{b}\right)\frac{\sum_{u=1}^N\hat\eta^2_u}{\sum_{u=1}^N\hat\eta_u} \\
    & \le
    \frac{D_{\cal L}^2}{N \eta_1} + \frac{D_{\cal L}^2}{N\psi\eta_2} + \left(\sigma^2 + 3\frac{\gamma^2}{b}\right) {B} \psi \eta_2 \\
    & \le
    \frac{\beta B D_{\cal L}^2}{N} + \sqrt{\frac{B\sigma^2 + 3B\gamma^2/b}{N}} \left(\frac{{\cal D}_{\cal L}^2}{\psi} + \psi\right)
    \le
    \frac{\beta B D_{\cal L}^2}{N} + 2{\cal D}_{\cal L}\sqrt{\frac{B\sigma^2 + 3B\gamma^2/b}{N}}
\end{align*}
To conclude the proof it remains to provide a bound in the case of convex loss function. Due to the smoothness of the loss function:
\begin{gather}\label{eq:smooth}
\frac{1}{\beta} \|\nabla {\cal L}(\omega_u)\|_2^2 \le \langle \nabla {\cal L}(\omega_u), \omega_u - \omega_*\rangle
\end{gather}
Denote $\phi_u = \omega_u^0 - \omega_*$, then
\begin{align*}
\phi_{u+1}^{2} & = \left\|\omega_{u+1}^2 - \eta_u \sum_{t\in {\cal B}_u} g_u^t  - \omega_*\right\|_2^2 \\
\phi_{u+1}^{2}  & = \phi_u^2 - 2\eta_u \sum_{t \in {\cal B}_u}\langle\delta_u^t, \omega_u - \omega_*\rangle + \eta_u^2 \left\| \sum_{t\in {\cal B}_u} g_u^t\right\|_2^2\\
               & = \phi_u^2  - 2\eta_u \sum_{t\in {\cal B}_u}\langle \nabla {\cal L}(\omega_u) + \delta_u^t, \omega_u - \omega^*\rangle  \\
               & \hspace{20mm} + \eta_u^2 \left(\|\nabla {\cal L}(\omega_u)\|_2^2 + 2\sum_{t \in {\cal B}_u}\langle\nabla {\cal L}(\omega_u), \delta_u^t\rangle + \left\|\sum_{t\in {\cal B}_u}\delta_u^t\right\|_2^2\right)
\end{align*}
Combining it with the smoothness condition, Eq.~\eqref{eq:smooth}, we have
\begin{align}\label{eq:conv_th1}
\phi_{u+1}^2 - \phi_u^2 \le - (2|{\cal B}_u|\eta_u - \beta|{\cal B}_u|^2 \eta_u^2)&[{\cal L}(\omega_u) - {\cal L}(\omega_*)] \nonumber\\
                            & \hspace{-10mm} - 2\eta_u\sum_{t\in {\cal B}_u} \langle\omega_k - \omega_* - \eta_u \nabla f(\omega_u), \delta_u^t\rangle
                            + \eta_u^2 \left\|\sum_{t\in {\cal B}_u}\delta_u^t\right\|_2^2
\end{align}
Summing up the Inequalities~\eqref{eq:conv_th1} above for all $u$, we have
\begin{align*}
  & \sum_{u=1}^N \left(\hat\eta_u - \frac{\beta}{2}{\hat\eta}^2_u\right) ({\cal L}(\omega_u) - {\cal L}(\omega_*)) \\
  & \hspace{20mm}\le {\cal D}_{\omega}^2 - 2 \sum_{u=1}^N \sum_{t \in {\cal B}_u}\eta_u \langle\omega_u - \eta_u \nabla {\cal L}(\omega_u) - \omega^*, \delta_u^t\rangle + \sum_{u=1}^N\eta_u\left\|\sum_{t \in {\cal B}_u} \delta_u^t\right\|_2^2
\end{align*}
The rest of the proof exactly follow along the lines of that of first part and hence the details are omitted.
\end{Proof}

\end{document}